\documentclass[aps,
prd,
notitlepage,twocolumn,showpacs,superscriptaddress,nofootinbib,preprintnumbers]{revtex4-2}  
\usepackage{graphicx}  
\usepackage{dcolumn}   
\usepackage{bm}        
\usepackage{amssymb}   
\usepackage{amsmath}
\usepackage{physics}
\usepackage{slashed}
\usepackage{mathtools}
\usepackage[colorlinks=true,citecolor=teal,urlcolor=teal,linkcolor=purple]{hyperref}
\usepackage{cleveref}
\usepackage{orcidlink}
\usepackage{tikz}
\usepackage{subcaption}
\usetikzlibrary{decorations.pathmorphing}
\usetikzlibrary{decorations.markings}
\usetikzlibrary{positioning, shapes, snakes, arrows}
\usetikzlibrary{patterns.meta}
\usetikzlibrary{arrows.meta}

\usepackage{soul}
\usepackage[normalem]{ulem}

\usepackage{tikz}
\usetikzlibrary{decorations.pathmorphing}
\usetikzlibrary{decorations.markings}
\usetikzlibrary{positioning, shapes, decorations, arrows}
\usetikzlibrary{patterns}

\tikzset{
  externalShorter/.style={
    shorten <=2mm,
    shorten >=2mm
  },
  wl/.style={line width=1pt},
  graviton/.style={line width=.8pt, -latex,decorate, decoration={snake, segment length=4pt,amplitude=1.8pt, pre length=.15cm, post length=.25cm}},
    worldlineStatic/.style={dotted, line width=1pt},
	worldline/.style={gray, line width=1pt},
	worldlineBold/.style={black, line width=.6pt},
	zUndirected/.style={line width=1pt},
	zParticle/.style={line width=1pt,postaction={decorate},decoration={markings,mark=at position .6 with {\arrow[#1]{latex}}}},
	zParticleTest/.style={white,line width=1pt,postaction={decorate},decoration={markings,mark=at position .6 with {\arrow[#1]{latex}}}},
  example1/.style={
    draw=none,
    postaction={decorate},
    decoration={
      markings,
      mark=at position #1 with {
        \arrow[
          black,
          scale=2.4
        ]{Straight Barb}
      }
    }
  },
  example2/.style={
    white,
    line width=1pt,
    postaction={decorate},
    decoration={
      markings,
      mark=at position .6 with {
        \arrow[
          draw=black,
          scale=1.5,
          #1
          ]{latex}
      }
    }
  },
	zParticle2/.style={line width=1pt,postaction={decorate},decoration={markings,mark=at position .75 with {\arrow[#1]{latex}}}},
	zParticle1/.style={line width=1pt},
	worldlineCut/.style={dotted,line width=1pt,postaction={decorate},decoration={markings,mark=at position .7 with {\arrow[#1]{latex}}}},
	worldlineCut2/.style={dotted,line width=1pt,postaction={decorate},decoration={markings,mark=at position .6 with {\arrow[#1]{latex}}}},
	zParticleF/.style={line width=1pt,postaction={decorate}},
	cscalar/.style={line width=1pt,postaction={decorate},decoration={markings,mark=at position .6 with {\arrow[#1]{latex}}}},
	cscalar2/.style={line width=1pt,postaction={decorate},decoration={markings,mark=at position .8 with {\arrow[#1]{latex}}}},
	photon/.style={line width =.8pt, decorate, decoration={snake, segment length=4pt, amplitude=1.8pt,  pre length=.1cm, post length=.1cm}},
	photonTest/.style={
    line width =.8pt,
    decorate,
    decoration={
      snake,
      segment length=5pt,
      amplitude=1.4pt,
      pre length=.0cm,
      post length=.0cm}},
	photonRed/.style={red, line width =.8pt, decorate, decoration={snake, segment length=4pt, amplitude=1.8pt,  pre length=.1cm, post length=.1cm}},
	cross/.style={cross out, line width =.8pt, draw=black, minimum size=2*(#1-\pgflinewidth), inner sep=0pt, outer sep=0pt},
cross/.default={4pt},
  dottedLine/.style={
    dotted, thick
  }
}
\usetikzlibrary{calc}

\newcommand{\drawLtoLshort}{
  \begin{scope}[shift={(currentLocation)}]
    \draw[photonTest,red] (0,.8) -- (.6,.8) ;
    \draw[dottedLine] (0,0) -- (.6,0) ;
  \end{scope}
  \coordinate (currentLocation) at ($(currentLocation)+(.6,0)$) ;
}
\newcommand{\drawLtoLmedium}{
  \begin{scope}[shift={(currentLocation)}]
    \draw[photonTest,red] (0,.8) -- (.8,.8) ;
    \draw[dottedLine] (0,0) -- (.8,0) ;
  \end{scope}
  \coordinate (currentLocation) at ($(currentLocation)+(.8,0)$) ;
}
\newcommand{\drawLzero}{
  \begin{scope}[shift={(currentLocation)}]
    \draw[photonTest] (0,0) -- (0,.8) ;
    \vertexSmall{(0,0)}
    \vertexSmall{(0,.8)}
  \end{scope}
}
\newcommand{\drawLoneA}{
  \begin{scope}[shift={(currentLocation)}]
    \draw[photonTest] (.3,0) -- (0,.8) ;
    \draw[photonTest] (-.3,0) -- (0,.8) ;
    \vertexSmall{(.3,0)}
    \vertexSmall{(-.3,0)}
    \vertexSmall{(0,.8)}
  \end{scope}
}

\newcommand{\drawLtwoA}{
  \begin{scope}[shift={(currentLocation)}]
    \draw[photonTest] (.4,0) -- (0,.8) ;
    \draw[photonTest] (-.4,0) -- (0,.8) ;
    \draw[photonTest] (0,0) -- (0,.8) ;
    \vertexSmall{(.4,0)}
    \vertexSmall{(0,0)}
    \vertexSmall{(-.4,0)}
    \vertexSmall{(0,.8)}
  \end{scope}
}
\newcommand{\drawLthreeA}{
  \begin{scope}[shift={(currentLocation)}]
    \draw[photonTest] (.6,0) -- (0,.8) ;
    \draw[photonTest] (-.6,0) -- (0,.8) ;
    \draw[photonTest] (.2,0) -- (0,.8) ;
    \draw[photonTest] (-.2,0) -- (0,.8) ;
    \vertexSmall{(.6,0)}
    \vertexSmall{(.2,0)}
    \vertexSmall{(-.2,0)}
    \vertexSmall{(-.6,0)}
    \vertexSmall{(0,.8)}
  \end{scope}
}

\pgfdeclarelayer{foreground}
\pgfsetlayers{main,foreground}

\newcommand{\vertexSmall}[1]{
  \begin{pgfonlayer}{foreground}
    \draw[fill] #1 circle (.06);
    \end{pgfonlayer}
    }

\allowdisplaybreaks

\newcommand{\cO}{\mathcal{O}}
\newcommand{\cM}{\mathcal{M}}

\newcommand{\vareps}{\varepsilon}

\newcommand{\iO}{i 0^{+}}
\newcommand{\ePM}{\epsilon_{\rm PM}}

\newcommand{\kays}[1]{\textbf{\textcolor{orange}{[Kays: #1]}}}
\newcommand{\fabian}[1]{\textbf{\textcolor{green}{[Fabian: #1]}}}

\begin{document}


\title{Gravitational wave scattering at $\cO(G^4)$: Murua construction and elliptics}

\author{Yilber Fabian Bautista\,\orcidlink{0000-0001-6255-5675}}
\affiliation{Higgs Centre for Theoretical Physics, School of Physics and Astronomy, The University of Edinburgh, \\
Edinburgh EH9 3JZ, Scotland, UK}
\author{Mathias Driesse\,\orcidlink{0000-0002-3983-5852}}
\affiliation{Institut f\"ur Physik, Humboldt-Universit\"at zu Berlin,
 10099 Berlin, Germany}
 \author{Kays Haddad\,\orcidlink{0000-0002-1182-2750}}
\affiliation{Institut f\"ur Physik, Humboldt-Universit\"at zu Berlin,
 10099 Berlin, Germany}
\author{Gustav Uhre Jakobsen\,\orcidlink{0000-0001-9743-0442}}
\affiliation{Institut f\"ur Physik, Humboldt-Universit\"at zu Berlin,
 10099 Berlin, Germany}

\date{\today}

\begin{abstract}
We compute the amplitude for the scattering of a gravitational wave off of a spinless point particle at fourth order in Newton's constant, using the worldline quantum field theory formalism.
A decomposition of our master integrals incorporating Murua coefficients allows us to entirely bypass the cut subtraction needed to convert the scattering amplitude into the Magnusian, the latter mapping directly onto the scattering phase shift in partial wave space.
Elliptic functions appear in momentum space for the first time for this process at this order.
To compare to the phase shift from black hole perturbation theory, we present an analytical method for performing the partial wave projection of the new elliptic sector to any partial wave order $\ell$, and verify the equivalence of the Magnusian and the phase shift.
This demonstrates that point-particle worldline quantum field theory accurately describes Schwarzschild black holes up to $\cO(G^4)$.
\end{abstract}

\pacs{}

\preprint{HU-EP-26/19}

\maketitle


\section{Introduction}

Due to their demand for highly accurate waveform templates, now-routine and upcoming higher-sensitivity gravitational wave measurements challenge the theoretical modeling of compact binary coalescence in general relativity (GR) \cite{KAGRA:2013rdx,LISA:2024hlh}.
Effective field theoretic (EFT) techniques have risen to the challenge, importing a rich catalogue of tools from quantum field theory to this classical setup \cite{Goldberger:2004jt,Porto:2005ac,Levi:2015msa,Porto:2016pyg,Levi:2018nxp,Damgaard:2019lfh,Kalin:2020fhe,Jakobsen:2021zvh,Bern:2020buy,Levi:2020uwu,Goldberger:2022ebt,Ben-Shahar:2023djm,Haddad:2024ebn}.
Such frameworks model compact bodies as point particles corrected with higher-dimensional operators, organized into powers of the curvature tensor.
This paradigm must be validated by matching EFT predictions to GR solutions, as has been done to establish a set of linear-in-curvature operators faithfully describing both Schwarzschild and Kerr black holes (BHs) \cite{Levi:2015msa,Vines:2017hyw}.

Wilsonian naturalness anticipates that quadratic-in-curvature operators are first relevant to binary dynamics at the sixth post-Minkowskian (6PM, $\cO(G^{6})$) order \cite{Goldberger:2004jt,Porto:2005ac,Hinderer:2007mb,Damour:2009vw,Levi:2015msa,Levi:2018nxp,Levi:2020uwu,Goldberger:2022ebt,Haddad:2024ebn}; their associated Wilson coefficients must also be fixed to BH values by matching to GR.
While originally motivated by the modeling of Kerr black holes as elementary spinning particles \cite{Guevara:2018wpp,Chung:2018kqs,Bautista:2021wfy,Bern:2022kto,Aoude:2022trd,Bautista:2022wjf,Bautista:2023szu,Bautista:2023sdf,Akpinar:2025byi,Akpinar:2025huz}, the nearly complete 5PM binary scattering dynamics \cite{Driesse:2024xad,Driesse:2024feo,Bern:2025wyd,Driesse:2026qiz} give this matching new urgency in the absence of spin; in this regard, various approaches have demonstrated the vanishing of the first layer of Love numbers -- the static Love numbers -- for black holes \cite{Kol:2011vg,Parra-Martinez:2025bcu}.

Gravitational wave scattering off of a BH background is a natural arena for testing quadratic-in-curvature augmentations to the EFT.
In addition to the tidal effects encoded by these operators, the GR description of this process -- elaborated by BH perturbation theory (BHPT) \cite{Matzner:1977,Matzner:1978} -- contains information about long-range gravitational exchanges.
One must therefore include this information in an EFT matching as well, involving technically demanding multi-loop calculations.

Fortunately, the extensive machinery developed for the two-body problem can be adapted to this setting \cite{Bern:2019nnu,Kalin:2020mvi,Kalin:2020fhe,Bern:2021dqo,Dlapa:2021npj,Dlapa:2021vgp,Bern:2021yeh,Kalin:2022hph,Jakobsen:2022fcj,Jakobsen:2023ndj,Dlapa:2023hsl,Driesse:2024xad,Driesse:2024feo,Bern:2025wyd,Driesse:2026qiz,Bohnenblust:2026ujk}.
In this vein, we developed in ref.~\cite{Bautista:2026qse} the framework for matching worldline QFT (WQFT) \cite{Mogull:2020sak} with BHPT through the scattering amplitude for the aforementioned process.
Carrying out the matching at the level of the elastic phase shift, we demonstrated that point particles and Schwarzschild BHs are equivalent up to next-to-next-to-leading order, $\cO(G^3)$; see also \cite{Bjerrum-Bohr:2025bqg,Ivanov:2026icp,Bjerrum-Bohr:2026fhs}.

Here we advance this program to the next order in Newton's constant, entailing a three-loop calculation.
We explicitly show that, after projecting onto a spherical wave basis, WQFT reproduces the corresponding BHPT phase shift.
As corrections to the spinless point-particle description are not yet relevant at this precision, this establishes the indistinguishability of Schwarzschild black holes from point particles up to $\cO(G^4)$.

Our work here provides an opportunity to probe additional aspects of the calculation.
First, incorporating the so-called Murua coefficients \cite{Kim:2024svw,Brandhuber:2025igz,Gonzo:2026yha}, we implement the conversion between the scattering amplitude and the Magnusian \cite{Damgaard:2021ipf,Kim:2024svw,Kim:2025gis} directly at the level of the master integrals, a drastic improvement over ref.~\cite{Bautista:2026qse}.
Second, we find that the infrared finiteness of the Magnusian persists at three-loop order.
Finally, we observe that the WQFT amplitude at $\mathcal{O}(G^4)$ displays a richer-than-anticipated mathematical structure, with elliptic functions appearing for the first time.

Complementary techniques employing the background field method \cite{Ivanov:2024sds}, an effective wave equation~\cite{Correia:2024jgr,Caron-Huot:2025tlq}, and the formulation of a shell-EFT \cite{Kosmopoulos:2025rfj} have recently been developed for scalar-wave scattering.
Extensions to waves of other spin weights have likewise appeared \cite{Combaluzier--Szteinsznaider:2025eoc,Solon:2026ubm}.
Insights from wave scattering off BHs have also found applications to BH binaries in the identification of universal contributions to the gravitational waveform \cite{Ivanov:2025ozg,Cipriani:2026xmx,Chang:2026eti}.

Even in view of this remarkable progress, WQFT sets itself apart in the ease of incorporating spin effects \cite{Jakobsen:2021zvh,Haddad:2024ebn}.
Ultimately, we aim to pin down an EFT for Kerr BHs beyond linear order in curvature and fourth order in spin (see e.g. refs.~\cite{Bern:2022kto,Bonocore:2025stf,Ben-Shahar:2025tiz} for non-BHPT-based approaches).

We begin the rest of this paper with a recap of the procedure developed in ref.~\cite{Bautista:2026qse} for matching WQFT to BHPT.
Then, in \Cref{sec:Murua}, we discuss how to leverage the Murua coefficients to easily extract the Magnusian from the amplitude.
Feynman integration is addressed in \Cref{sec:Integration}.
Finally, the result of our computation and matching to BHPT can be found in \Cref{sec:Results}.

\section{Matching WQFT to BHPT}

In WQFT, a gravitating point particle with mass $m$ and without spin is described by the action \cite{Mogull:2020sak}
\begin{align}\label{eq:WQFT}
  S_{\rm wl}=-\frac{m}{2}\int{\rm d}\tau\,g_{\mu\nu}(x)\dot{x}^\mu(\tau)\dot{x}^\nu(\tau).
\end{align}
Graviton dynamics are given by the usual Einstein-Hilbert action in conjunction with the gauge of ref.~\cite{Driesse:2024xad}, preserving the de Donder propagator while simplifying graviton vertices.
Particle dynamics are restricted by \cref{eq:WQFT} to a one-dimensional worldline, whereas pure-graviton interactions take place in a $d$-dimensional bulk, taking $d=4-2\varepsilon$ for the purposes of dimensional regularization.

On the WQFT side, the amplitude we are after is the $S$-matrix element with an initial graviton state with momentum $k_{1}^{\mu}$ and helicity $h_{1}$ and a final graviton with momentum $k_{2}^\mu$ and helicity $h_{2}$.
Diagrammatically,
\begin{equation}
\begin{aligned}\label{eq:DefTMatrixElement}
  \langle k_2,h_2|i\hat T| k_1,h_1\rangle
  \,\,&=\,\, 
  \begin{tikzpicture}[baseline=4mm,scale = 0.75]
    \draw [dotted, thick] (-1.5,0) -- (1.5,0) ;
    \draw [photonTest] (0,0) -- (-1.2,1.2) ;
    \draw [example1={.5}]  (-1.2,1.2) -- (0,0) ;
    \draw [photonTest] (0,0) -- (1.2,1.2) ;
    \draw [example1={.7}] (0,0) -- (1.2,1.2) ;
    \node[right] at (0.75,1.45) {$k_2,h_2$} ;
    \node[left] at (-0.75,1.45) {$k_1,h_1$} ;
    \draw [fill=black!60, thick] (0,0.1) circle (0.4cm);
  \end{tikzpicture} \\
  &=\,\, 
  2\pi\,\delta(k_2\cdot v-k_1\cdot v)
  \,i
  \cM_{\sigma}(\theta,\phi),
\end{aligned}
\end{equation}
where $\hat{S}=1+i\hat{T}$.
We adopt the kinematics and notation of ref.~\cite{Bautista:2026qse}.
Namely, the incoming and outgoing graviton momenta and polarizations are
\begin{equation}\label{eq:Kinematics}
    \begin{aligned}
        k_{1}^{\mu}
        &=
        \omega(1,0,0,1)
		,\quad
		\epsilon_{1}^{\sigma_{1},\mu}
        =
        \frac{1}{\sqrt{2}}(0,1,i\sigma_{1},0) \\
        k_{2}^{\mu}
        &=
        \omega(1,s_\theta\,c_\phi,s_\theta\,s_\phi,c_\theta), 
        \\ 
        \epsilon_{2}^{\sigma_{2},\mu}
        &=
        \frac{1}{\sqrt{2}}\left(
          0,c_\theta\,c_\phi-i\sigma_{2}s_\phi,c_\theta\,s_\phi+i\sigma_{2}c_\phi,-s_\theta
          \right),
    \end{aligned}
\end{equation}
where $s_{y}\equiv\sin y$ and $c_{y}\equiv\cos y$.
Additionally, graviton helicities are denoted by their signs $\sigma_{i}$, such that $h_{i}=2\sigma_{i}$.
When only the product of the helicities, rather than their individual values, is relevant, we write $\sigma\equiv\sigma_{1}\sigma_{2}$.
As $k_{2}^{\mu}$ is outgoing, it is the conjugate of its polarization which enters the amplitude.
Finally, we work in the rest frame of the massive body, $v^{\mu}=(1,\boldsymbol{0}) $, implying the scattering is a non-trivial function only of the scattering angle $\theta$, encoded in the variable $x\equiv\sin(\theta/2)$.

To compute the amplitude in \cref{eq:DefTMatrixElement} we resort to a perturbative expansion around flat space, where the metric and worldline trajectory are expanded according to
\begin{equation}
\begin{aligned}
  g_{\mu\nu}(x)&=\eta_{\mu\nu}+\kappa\,h_{\mu\nu}(x), \\
  x^{\mu}(\tau)&=v^\mu\tau+z^\mu(\tau),
\end{aligned}
\end{equation}
having centered our coordinate system on the particle's position at $\tau=0$.
The perturbative parameter is $\kappa=\sqrt{32\pi G}$, such that the trajectory perturbation $z^\mu(\tau)\sim\cO(\kappa^{2})$.
In the remainder of this paper, we organize perturbative expansions in powers of $\ePM\equiv2Gm\omega$.

Within the BHPT framework, this amplitude is extracted from the solution for curvature perturbations of a Schwarzschild spacetime; see e.g. ref.~\cite{Futterman_Handler_Matzner_1988}.
Written in a spherical wave basis, the helicity-preserving ($\sigma=1$) and reversing ($\sigma=-1$) BHPT amplitudes are respectively
\begin{subequations}\label{eq:AmplitudeBHPT}
\begin{align}
  f(\theta,\phi) &=
  \frac{\pi}{\omega}\sum_{\ell=2}^\infty \sqrt{\frac{2\ell+1}{\pi}}{}_{-2}Y_{\ell 2}(\theta,\phi)\label{eq:ThpBHPT} \\
  &\times \sum_{P=\pm1} \left({}_{-2}\eta_{\ell2}e^{2i{}_{-2}\delta_{\ell 2}^{P}}-1\right)\,,\notag\\
  g(\theta,\phi) &=
  \frac{\pi}{\omega}\sum_{\ell=2}^\infty \sqrt{\frac{2\ell+1}{\pi}}{}_{-2}Y_{\ell 2}(\pi-\theta,\phi)\label{eq:ThrBHPT} \\
  &\times \sum_{P=\pm1} (-1)^\ell P\left({}_{-2}\eta_{\ell2}e^{2i{}^{\vphantom{P}}_{-2\vphantom{\ell}}\delta_{\ell 2}^{P}}-1\right)\,.\notag
\end{align}
\end{subequations}
All angular dependence of the scattering is described by the spin-weighted spherical harmonics ${}_{-2}Y_{\ell2}(\theta,\phi)$, with the scattering phase shift ${}_{-2}\delta^{P}_{\ell2}$ and absorption factor ${}_{-2}\eta_{\ell2}$ respectively encoding the remaining conservative and dissipative scattering information.
Similarly to the scattering amplitude, the latter two admit a perturbative expansion in Newton's constant, with dissipation only becoming relevant from $\cO(\ePM^5)$.

The amplitude derived from WQFT can be matched directly onto that from BHPT:
\begin{equation}\label{eq:MToBHPT}
\begin{aligned}
	\cM_{1,{\rm fin.}}(\theta,\phi)&=2\pi\, f(\theta,\phi) \\ \cM_{-1,{\rm fin.}}(\theta,\phi)&=2\pi\, g(\theta,\phi).
\end{aligned}
\end{equation}
However, this matching is subtle, as vanishing pieces of the amplitude in the $d\rightarrow4$ limit interfere with the resummed Weinberg phase to produce the finite amplitude $\cM_{\sigma,{\rm fin.}}$ needed for the matching.
A suitable prescription for matching at the level of the amplitude was put forth in ref.~\cite{Bjerrum-Bohr:2026fhs}.
While we do not check \cref{eq:MToBHPT} in this work, we discuss in \Cref{sec:Results} the construction of the finite amplitude which matches BHPT.

An alternative approach, requiring only up to the finite part of the amplitude in the $d\rightarrow 4$ limit, passes through the exponential representation of the $S$-matrix \cite{Damgaard:2021ipf},
\begin{align}
  \hat{S}=e^{i\hat{N}},
\end{align}
where $\hat{N}$ is referred to as the Magnus operator \cite{Kim:2024svw}.
As we demonstrated in ref.~\cite{Bautista:2026qse} (see also refs.~\cite{Ivanov:2024sds,Ivanov:2026icp}), the $N$-matrix element (or Magnusian \cite{Kim:2024svw,Kim:2025gis})
\begin{equation}
\begin{aligned}
  2\pi\, \delta(v\cdot k_{1}-v\cdot k_{2})&N_{h_{1},h_{2}}^{(n)}(\omega;\theta,\phi) \\
  &\equiv\langle k_{2},h_{2}|\hat{N}|k_{1},h_{1}\rangle|_{\cO(\epsilon_{\rm PM}^{n})}
\end{aligned}
\end{equation}
can be directly related to the elastic phase shift.
Specifically, at $O(\epsilon_{\text{PM}}^4)$
\begin{subequations}\label{eq:NToPhaseShift}
  \begin{align}
    \frac{1}{4\pi}\frac{N_{\ell;2,2}^{(4)}}{\sqrt{\pi(2\ell+1)}}&=\sum_{P=\pm}{}_{-2}\delta_{\ell 2}^{P,(4)},\label{eq:match} \\
    \frac{1}{4\pi}\frac{N_{\ell;2,-2}^{(4)}}{\sqrt{\pi(2\ell+1)}}&=\sum_{P=\pm}P(-1)^{\ell}\,{}_{-2}\delta_{\ell 2}^{P,(4)},
  \end{align}
\end{subequations}
where the left-hand side involves the projection of the $N$-matrix element on a basis of spin-weighted spherical harmonics,
\begin{subequations}\label{eq:deltalWQFT}
  \begin{align}
  N_{\ell;2,2}^{(4)}&=\omega\int{\rm d}\Omega\, {}_2 Y_{\ell, -2}(\theta,\phi) N^{(4)}_{2,2}(\omega;\theta,\phi) \,,\label{eq:deltalWQFThp}\\ 
  N_{\ell;2,-2}^{(4)}&=\omega\int{\rm d}\Omega\, {}_2 Y_{\ell,- 2}(\pi-\theta,\phi) N^{(4)}_{2,-2}(\omega;\theta,\phi)\label{eq:deltalWQFThr}\,.
  \end{align}
\end{subequations}
As we will see, the Magnusian for this process at the present order is infrared safe and finite in the forward limit $x\rightarrow 0$.
This justifies the above projection onto a basis of functions in four-dimensions;
see refs.~\cite{Ivanov:2024sds,Ivanov:2026icp} for a $d$-dimensional treatment.

\section{Efficient extraction of $N$}\label{sec:Murua}

\Cref{eq:NToPhaseShift} illustrates the suitability of $\hat{N}$ for matching to BHPT.
However, $\hat{T}$ is still crucial to the calculation as the perturbation theory is set up to produce its matrix elements (i.e. scattering amplitudes) rather than those of $\hat{N}$.
In ref.~\cite{Bautista:2026qse}, we converted the matrix elements of the former to those of the latter by subtracting cuts.
Here we take an alternative, much more efficient approach.

In the vein of the analysis in ref.~\cite{Haddad:2025cmw}, the effect of the aforementioned cut subtraction can be understood diagrammatically.
For example, at one-loop order, the $N$- and $T$-matrix elements are related by \cite{Damgaard:2021ipf}
\begin{align*}
  \begin{tikzpicture}[baseline=4mm,scale = 0.75]
    \draw [dotted, thick] (-1.5,0) -- (1.5,0) ;
    \draw [photon] (0,0) -- (-1.2,1.2) ;
    \draw [photon] (0,0) -- (1.2,1.2) ;
    \draw [fill=white, thick] (0,0.1) circle (0.4cm);
    \draw [pattern = north east lines, thick] (0,0.1) circle (0.4cm);
    \draw [fill=white, thick] (0,0.1) circle (0.2cm);
  \end{tikzpicture}
  \,\,
  &=
  \,\,
  \begin{tikzpicture}[baseline=4mm,scale = 0.75]
    \draw [dotted, thick] (-1.5,0) -- (1.5,0) ;
    \draw [photon] (0,0) -- (-1.2,1.2) ;
    \draw [photon] (0,0) -- (1.2,1.2) ;
    \draw [fill=black!60, thick] (0,0.1) circle (0.4cm);
    \draw [fill=white, thick] (0,0.1) circle (0.2cm);
  \end{tikzpicture} \\*
  &-\frac{1}{2}
  \,\,
  \begin{tikzpicture}[baseline=4mm,scale = 0.75]
    \draw [dotted, thick] (-1.5,0) -- (3,0) ;
    \draw [dashed, red, very thick] (0.75,1.25) -- (0.75,0.25) ;
    \draw [photon] (0,0) -- (-1.2,1.2) ;
    \draw [photon] (0,0) arc (180:0:.75);
    \draw [photon] (1.5,0) -- (2.7,1.2) ;
    \draw [fill=black!60, thick] (0,0.1) circle (0.4cm);
    \draw [fill=black!60, thick] (1.5,0.1) circle (0.4cm);
  \end{tikzpicture},
\end{align*}
with the striped one-loop blob representing the one-loop Magnusian, the solid one-loop blob the one-loop amplitude, and the red dashed line a cut of the exchanged graviton.
We may expand the cut into retarded and advanced propagators,\footnote{When the incoming graviton energy is positive, retarded propagators become Feynman propagators, and advanced become Dyson.}
\begin{align*}
  \begin{tikzpicture}[baseline=4mm,scale = 0.75]
    \draw [dotted, thick] (-1.5,0) -- (3,0) ;
    \draw [dashed, red, very thick] (0.75,1.25) -- (0.75,0.25) ;
    \draw [photon] (0,0) -- (-1.2,1.2) ;
    \draw [photon] (0,0) arc (180:0:.75);
    \draw [photon] (1.5,0) -- (2.7,1.2) ;
    \draw [fill=black!60, thick] (0,0.1) circle (0.4cm);
    \draw [fill=black!60, thick] (1.5,0.1) circle (0.4cm);
  \end{tikzpicture}
  \,\,
  &=i
  \,\,
  \begin{tikzpicture}[baseline=4mm,scale = 0.75]
    \draw [dotted, thick] (-1.5,0) -- (3,0) ;
    \draw [photon] (0,0) arc (180:0:.75);
    \draw [photon] (0,0) -- (-1.2,1.2) ;
    \draw [photon] (0,0) arc (180:0:.75);
    \draw [photon] (1.5,0) -- (2.7,1.2) ;
    \draw [fill=black!60, thick] (0,0.1) circle (0.4cm);
    \draw [fill=black!60, thick] (1.5,0.1) circle (0.4cm);
    \node at (0.75,0.85) [above] {$\rightarrow$};
  \end{tikzpicture} \\
  \,\,
  &-i
  \,\,
  \begin{tikzpicture}[baseline=4mm,scale = 0.75]
    \draw [dotted, thick] (-1.5,0) -- (3,0) ;
    \draw [photon] (0,0) -- (-1.2,1.2) ;
    \draw [photon] (0,0) arc (180:0:.75);
    \draw [photon] (1.5,0) -- (2.7,1.2) ;
    \draw [fill=black!60, thick] (0,0.1) circle (0.4cm);
    \draw [fill=black!60, thick] (1.5,0.1) circle (0.4cm);
    \node at (0.75,0.85) [above] {$\leftarrow$};
  \end{tikzpicture}
  \,\,,
\end{align*}
where the arrow pointing right indicates the retarded propagator.
This makes the function of the subtracted cut transparent: to modify the causality prescription of the active graviton, the internal graviton which may go on shell in the integration domain.
Therefore, in the one-loop case the result is simply to average over both prescriptions for the active graviton.
At higher loop orders averaging over both prescriptions is not the correct procedure, but a generalization does exist which bypasses the cut subtraction and produces the Magnusian directly.

Just as the traditional Dyson series underlies the computation of $T$-matrix elements, the alternative Magnus series \cite{Magnus1954} forms the foundation of $N$-matrix elements \cite{Kim:2024svw,Brandhuber:2025igz,Gonzo:2026yha}.
An essential difference is that, instead of the time-ordered products comprising the time-evolution operator in the former, the Magnus series involves commutators of the system's Hamiltonian.
Functionally, this induces diagrams with retarded and advanced -- rather than Feynman -- causality prescriptions.
Diagrams differing only by causality prescriptions receive relative weights determined by the so-called Murua coefficients.

At first, it may seem that we have dramatically increased the work to be done, as a diagram may possess many distinct prescriptions; however, we are spared by two considerations.
First, the Murua coefficients obey an edge-contraction rule -- described in ref.~\cite{Gonzo:2026yha} -- whose consequence is that distinct prescriptions must be considered only for active propagators.
For $L$-loop gravitational wave scattering, at most $2^{L}$ orientations are relevant.
Second, WQFT Feynman rules and integration-by-parts (IBP) reduction are agnostic to causality prescriptions, and vice versa.
This enables us to insert the relative weightings at the level of the master integrals.

Effectively, if an integral $I^{s_{1}\dots s_{n}}$ has $n$ propagators which may go on shell in the integration domain, we assign the integral a Murua value $I^{(M;n)}$ according to\footnote{Extrapolating from \cref{eq:MuruaIntegrals}, the Murua value of an integral with $n$ active propagators is the weighted sum over all possible causality prescriptions where an integral with $k$ advanced propagators receives the coefficient $[(n+1){{n}\choose{k}}]^{-1}$.
These weights sum to 1 and satisfy the edge-contraction rule.}
\begin{align}
  I^{(M;0)}&= I,\notag \\
  I^{(M;1)}&=\frac{1}{2}I^{+}+\frac{1}{2}I^{-},\notag \\
  I^{(M;2)}&=\frac{1}{3}I^{++}+\frac{1}{6}I^{+-}+\frac{1}{6}I^{-+}+\frac{1}{3}I^{--},\label{eq:MuruaIntegrals} \\
  I^{(M;3)}&=\frac{1}{4}I^{+++}+\frac{1}{12}I^{++-}+\frac{1}{12}I^{+-+}+\frac{1}{12}I^{-++}\notag \\
  &+\frac{1}{12}I^{+--}+\frac{1}{12}I^{-+-}+\frac{1}{12}I^{--+}+\frac{1}{4}I^{---}.\notag
\end{align}
On the right-hand side, the superscripts indicate whether each active propagator has a $+\iO$ or $-\iO$ prescription.
These relations imply that the Murua value of an integral is always real, and as a consequence, often has softer poles than the original integral, or even vanishes in many cases.

Summing up, then, the Magnusian we seek can be obtained by constructing the integrand with the usual Feynman rules, reducing to master integrals, and finally replacing each integral with its Murua value in \cref{eq:MuruaIntegrals}.
We emphasize again that the only difference between the amplitude and the Magnusian for this process is the causality prescription of the integrals.

\section{Integration}\label{sec:Integration}

Using the same recursive technique outlined in ref.~\cite{Bautista:2026qse}, we count 70 diagrams building up the integrand at three-loop order.
All Feynman integrals which appear can be brought into the three-loop family
\begin{align}\label{eq:intfamily3loops}
  I_{n_{1}\dots n_{12}}^{\eta_{1}\eta_{2}\eta_{3}}=\int_{\ell_{1},\ell_{2},\ell_{3}}\frac{1}{D_{1}^{n_{1}}\dots D_{12}^{n_{12}}},
\end{align}
having abbreviated the integration measures as
\begin{align*}
  \int_{\ell_{i}}=\tilde{\mu}^{2\varepsilon}\int\frac{{\rm d}^{d}\ell_{i}}{(2\pi)^{d}}2\pi\,\delta(v\cdot\ell_i-v\cdot k_{1}),
\end{align*}
with $\tilde{\mu}^{2}=e^{\gamma_{\rm E}}\mu^{2}/4\pi$ for some regularization scale $\mu$.
The propagators are
\begin{align*}
  D_{i=1,2,3}=\ell_{i}^{2}+\eta_{i}\iO,&\quad D_{i=4,5,6}=(\ell_{i-3}-k_{1})^{2} \\
  D_{i=7,8,9}=(\ell_{i-6}-k_{2})^{2},&\quad D_{i=10,11}=(\ell_{1}-\ell_{i-8})^{2} \\
  D_{i=12}&=(\ell_{2}-\ell_{3})^{2}.
\end{align*}
As is standard in multi-loop calculations, we adopt the method of differential equations to calculate our integrals.
Employing \texttt{Kira} \cite{Lange:2025fba} to generate IBP identities, we find a basis consisting of twenty master integrals when symmetries between integrals are not considered.

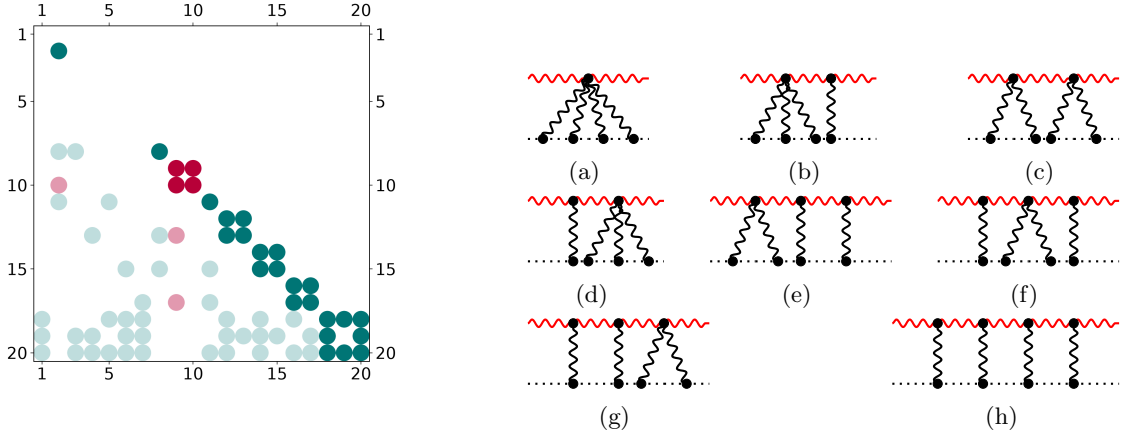
\begin{figure*}[!t]
  \centering
  \begin{subfigure}{.45\textwidth}
    \includegraphics[scale=0.3]{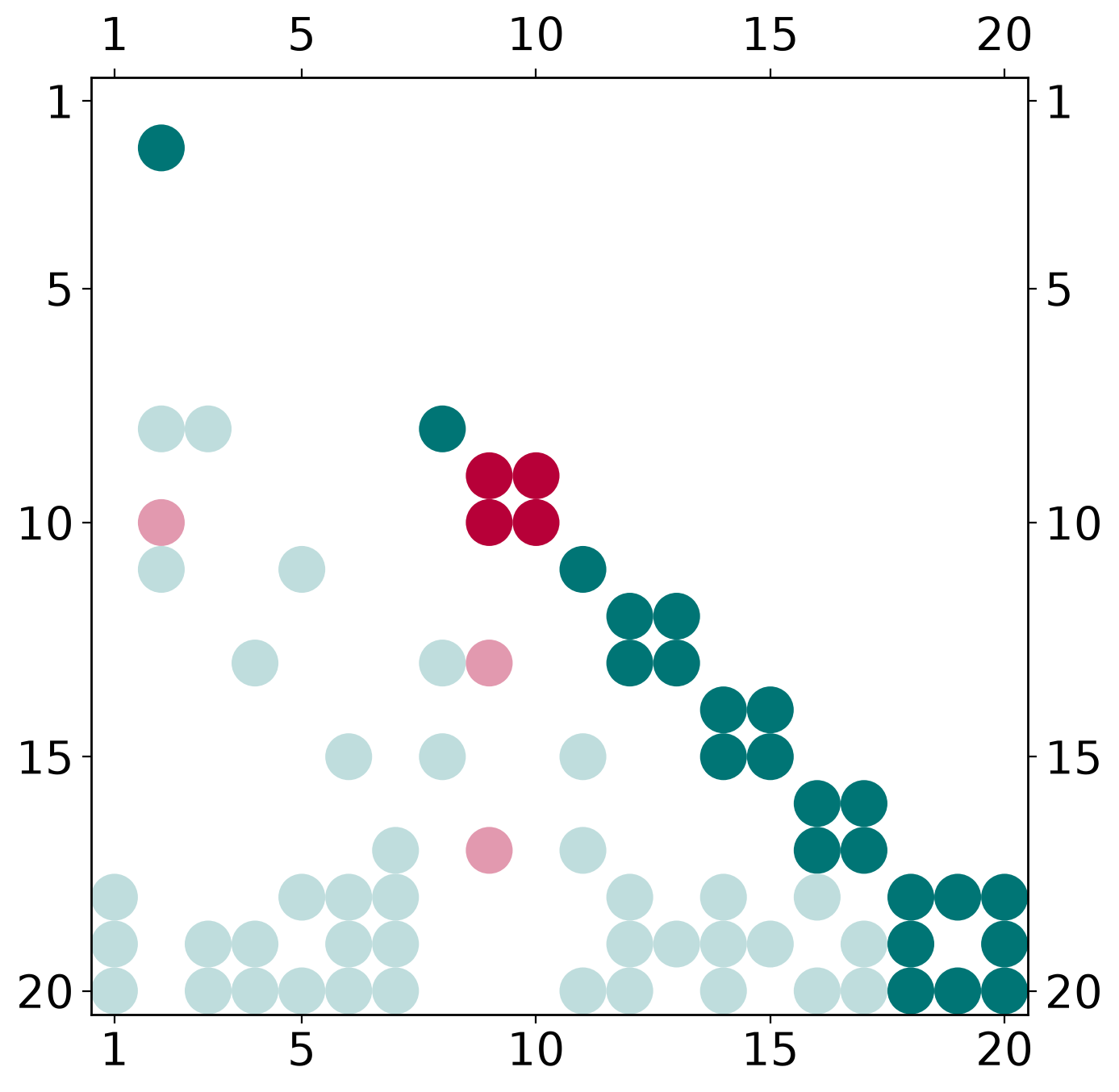}
  \end{subfigure}  
  \begin{subfigure}{.45\textwidth}
  \subcaptionbox{}{
    \begin{tikzpicture}
      \coordinate (currentLocation) at (0,0) ;
      \drawLtoLmedium
      \drawLthreeA
      \drawLtoLmedium
    \end{tikzpicture}
  }\hfill
  \subcaptionbox{}{
    \begin{tikzpicture}
      \coordinate (currentLocation) at (0,0) ;
      \drawLtoLshort
      \drawLtwoA
      \drawLtoLshort
      \drawLzero
      \drawLtoLshort
    \end{tikzpicture}
  }\hfill
  \subcaptionbox{}{
    \begin{tikzpicture}
      \coordinate (currentLocation) at (0,0) ;
      \drawLtoLshort
      \drawLoneA
      \drawLtoLmedium
      \drawLoneA
      \drawLtoLshort
    \end{tikzpicture}
  }\hfill
  \subcaptionbox{}{
    \begin{tikzpicture}
      \coordinate (currentLocation) at (0,0) ;
      \drawLtoLshort
      \drawLzero
      \drawLtoLshort
      \drawLtwoA
      \drawLtoLshort
    \end{tikzpicture}
  }\hfill
  \subcaptionbox{}{
    \begin{tikzpicture}
      \coordinate (currentLocation) at (0,0) ;
      \drawLtoLshort
      \drawLoneA
      \drawLtoLshort
      \drawLzero
      \drawLtoLshort
      \drawLzero
      \drawLtoLshort
    \end{tikzpicture}
  }\hfill
  \subcaptionbox{}{
    \begin{tikzpicture}
      \coordinate (currentLocation) at (0,0) ;
      \drawLtoLshort
      \drawLzero
      \drawLtoLshort
      \drawLoneA
      \drawLtoLshort
      \drawLzero
      \drawLtoLshort
    \end{tikzpicture}
  }\hfill
  \subcaptionbox{}{
    \begin{tikzpicture}
      \coordinate (currentLocation) at (0,0) ;
      \drawLtoLshort
      \drawLzero
      \drawLtoLshort
      \drawLzero
      \drawLtoLshort
      \drawLoneA
      \drawLtoLshort
    \end{tikzpicture}
  }\hfill
  \subcaptionbox{}{
    \begin{tikzpicture}
      \coordinate (currentLocation) at (0,0) ;
      \drawLtoLshort
      \drawLzero
      \drawLtoLshort
      \drawLzero
      \drawLtoLshort
      \drawLzero
      \drawLtoLshort
      \drawLzero
      \drawLtoLshort
    \end{tikzpicture}
  }
  \end{subfigure}
  \caption{
    \label{fig:matrix_plot}
    \textit{Left}: The canonicalized differential equation matrix $M_{\rm c}(x)$.
    Non-zero entries are represented by dots, with the elliptic block on the diagonal highlighted in red. 
    Elements which share a row or column with the elliptic block are in pink.
    All other elements only contain the letters $\{-1,0,1\}$.
    \textit{Right}: Integral topologies of the non-constant sectors of the differential equation, read down the diagonal in the plot on the left. Integral (c) introduces the elliptics.
  }
\end{figure*}

Given the IBP relations, a differential equation for the master integrals can be set up and brought to the canonical form
\begin{align}
  \frac{\rm d}{{\rm d}x}\vec{I}_{\rm c}(x;\varepsilon)=\varepsilon M_{\rm c}(x)\vec{I}_{\rm c}(x;\varepsilon).
\end{align}
Crucially, at this loop order, canonicalizing produces an elliptic sector with the associated Picard-Fuchs equation
\begin{equation}\label{eq:picard_fuchs}
  \mathcal{L}(\varpi(x))=\left[\frac{d}{dx}\left(x(1-x^2)\frac{d}{dx}\right)-x\right]\varpi(x)=0
\end{equation}
see \cref{fig:matrix_plot}. 
This is solved by the complete elliptic integral of the first kind,\footnote{Elliptic integrals have also been found in gravitationally-mediated scalar wave scattering, as presented by Giulia Isabella at the \href{https://www.ias.edu/video/gravitational-tidal-matching-and-elliptic-curves}{Black Holes from Theory to Observations workshop}.}
\begin{align}
  \varpi(x)=c_{1} K(x^2)+c_2 K(1-x^2).
\end{align}
We choose the $\{c_{1},c_{2}\}=\{1,0\}$ solution such that the differential equation is smooth in the vicinity of $x=0$ where we solve it.

Once canonicalized, the solution to the differential equation may be written as a path-ordered exponential
\begin{align}\label{eq:PathOrderExp}
  \vec{I}_{\rm c}(x;\varepsilon) = \mathcal{P}\exp\left[\varepsilon\int_{x_{0}}^{x}M_{\rm c}(x^{\prime}){\rm d}x^{\prime}\right]\vec{J}(\varepsilon),
\end{align}
and the problem is reduced to determining the boundary constants $\vec{J}(\varepsilon)$, which are independent of $x$; these are related to boundary integrals that encode the data on the various causality prescriptions of the master integrals.
Given these, \cref{eq:PathOrderExp} can be expanded to extract the canonical master integrals to the desired order in $\varepsilon$, expressed in terms of iterated integrals over $M_{\rm c}$.

To compute the boundary integrals, we expand \cref{eq:PathOrderExp} near $x=0$.
As at lower loop orders, all integrals in the vicinity of this point possess only two regions: the general region, where all loop momenta $\ell_{i}\sim 1\gg x$, giving integrals an $x^0$ scaling; and the forward region where all $\ell_{i}\sim x$, meaning the integrals scale as $x^{n-6\varepsilon}$, where $n\in\mathbb{Z}$ and the $-6\varepsilon$ power comes from the integral measure.

Not counting various $\iO$ prescriptions, a total of nine and six boundary integrals in each region respectively are sufficient to uniquely solve the differential equation.
We employ various techniques to determine these, including loop-by-loop integration, parametric representations, and reconstruction from high-precision numerical evaluation using \texttt{AMFlow} \cite{Liu:2022chg}.
Boundaries determined through numerics were evaluated up to 150 digits and fit to a basis comprising the transcendental constants $\{\pi,\log 2,\zeta_3\}$ and their products.
The uniqueness of the reconstruction was inferred from its stability over a wide range of fitted digits, and by overdetermination tests involving higher-transcendentality (products of the above) constants.
Complex conjugation, cut relations, and partial-fraction identities for forward-region integrals (see ref.~\cite{Driesse:2024xad}) give the remaining $\iO$ routings.

The persistence of only two regions in our integrals sets this problem apart from the binary-scattering computation.
There, while each loop momentum is also allowed only two distinct scalings, the loop momenta are not constrained to all scale the same way.
This produces three regions at 4PM \cite{Bern:2021yeh}, while four are needed in a putative conservative sector at 5PM-1SF \cite{Driesse:2024xad}.

\section{Gravitational wave scattering}\label{sec:Results}

The Murua prescription \cref{eq:MuruaIntegrals} directly produces the Magnusian from our WQFT integrand, which we may write in a PM decomposition:
\begin{align}\label{eq:Nhelicity}
  N_{h_{1},h_{2}}=\frac{e^{ih_{1}\phi}}{\omega}\sum_{n}(\pi\epsilon_{\rm PM})^{n}N^{(n)}_{\sigma}(x).
\end{align}
With the $n=1,2,3$ terms calculated in ref.~\cite{Bautista:2026qse}, we present here only the $n=4$ case.
In the helicity-reversing and preserving configurations, we find respectively 
\begin{widetext}
\begin{equation}\label{eq:3LoopNMatrix}
\begin{aligned}
  N^{(4)}_{-1}(x)&=0 \\
  N^{(4)}_{1}(x)&=\frac{1575 x^6+47655 x^5+827547 x^4+366747 x^3+3142262 x^2+166262 x+722432}{23040\pi^{2}(x+1)^{2}(x-1)} \\
  &
  +\frac{4(7+11x^2)}{\pi^2(1-x^2)}G(-1;x)
  -\left[1-\frac{12}{\pi^{2}}\,G(0,-1;x)\right]\frac{4}{3}\frac{1+6x^{2}+2x^{4}}{(x^{2}-1)^{2}} \\
  &-\frac{17805 x^4+113738 x^2+44273}{768 \pi ^2 (x^{2}-1)^{2}}\left[\frac{\mathcal{I}_{1}(x)}{K(x^{2})}-E(x^{2})\mathcal{I}_{2}(x)\right] +\frac{10125 x^4+107121 x^2+58570}{1536 \pi ^2(x^{2}-1)}K(x^{2})\mathcal{I}_{2}(x),
\end{aligned}
\end{equation}
\end{widetext}
where $G(-1;x)$ and $G(0,-1;x)$ are multiple polylogarithms, $K$ and $E$ are complete elliptic integrals of the first and second kinds, and $\mathcal{I}_{1,2}(x)$ are the iterated integrals
\begin{align}
  \mathcal{I}_{1}(x)&\equiv\int_{0}^{x}{\rm d}y\,K(y^{2}), \\
  \mathcal{I}_{2}(x)&\equiv\frac{\pi}{2}-\int_{0}^{x}{\rm d}y\,\frac{\mathcal{I}_{1}(y)}{y(1-y^{2})K^{2}(y^{2})}.
\end{align}
The first of these can be evaluated to a hypergeometric function.
\Cref{eq:3LoopNMatrix} extends to three-loop order the observation made in ref.~\cite{Bautista:2026qse} that the Magnusian is an infrared-safe quantity encoding the scattering process.

\begin{figure*}
  \begin{subfigure}{.45\textwidth}
    \centering
    \includegraphics[width=0.83\textwidth]{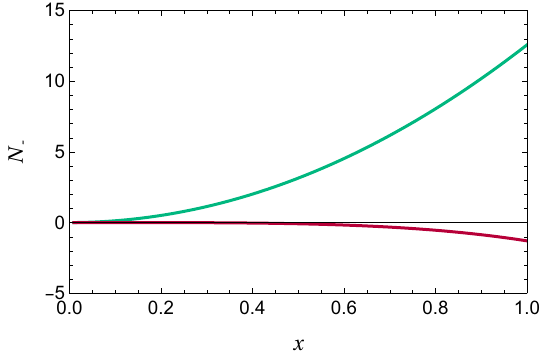}
    \caption{}
    \label{fig:myplot1}
  \end{subfigure}
%
%
  \begin{subfigure}{.45\textwidth}
    \centering
    \includegraphics[width=\textwidth]{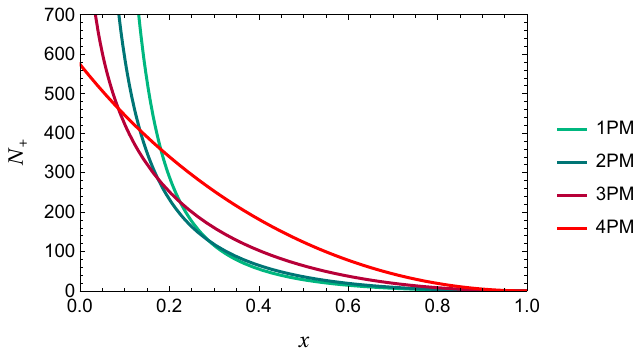}
    \caption{}
    \label{fig:myplot2}
  \end{subfigure}
  \caption{The (a) helicity-reversing and (b) preserving Magnusian $(\pi\ePM)^{n} N^{(n)}_{\sigma}(x)$, with kinematics such that $\ePM=2$ -- this non-perturbative value makes visible the behavior at higher PM orders.
  The helicity-reversing Magnusian is zero at 2 and 4PM, and is not plotted at these orders.
  Qualitatively, both helicity configurations vanish in opposite limits, and $N_{\sigma}$ becomes free of poles at 4PM; see \cref{eq:divF}.}
  \label{fig:Nofx}
\end{figure*}

It is immediate to take the forward limit of $N^{(4)}_{\sigma}$, producing a finite result.
In the backward limit, $x\rightarrow1$, the Magnusian must be smooth, as it possesses the same kinematic poles as the amplitude.
To facilitate this limit, note that the denominator of $\mathcal{I}_{2}(x)$ can be written as
\begin{align*}
  \frac{1}{x(1-x^{2})K^{2}(x^{2})}=-\frac{2}{\pi}\frac{\rm d}{{\rm d}x}\frac{K(1-x^{2})}{K(x^{2})},
\end{align*}
allowing us to integrate by parts:
\begin{equation}
\begin{aligned}\label{eq:I2sim}
  \mathcal{I}_{2}(x)=\frac{\pi}{2}&+\frac{2}{\pi}\mathcal{I}_{1}(x)\frac{K(1-x^{2})}{K(x^{2})} \\
  &-\frac{2}{\pi}\int_{0}^{x}{\rm d}y\,K(1-y^{2}).
\end{aligned}
\end{equation}
Its Taylor expansion in the vicinity of $x=1$ is now readily evaluated, giving
\begin{equation}
\begin{aligned}
  \mathcal{I}_{2}(x)&=-\frac{2}{\log \left(\frac{1-x}{8}\right)} \\
  &\quad-(1-x)\frac{7 \log\left(\frac{1-x}{8}\right)-9}{9 \log ^2\left(\frac{1-x}{8}\right)}{+}\cO((1-x)^{2}),
\end{aligned}
\end{equation}
which renders the $x\rightarrow1$ pole in \cref{eq:3LoopNMatrix} spurious.
The Magnusian is plotted as a function of $x$ in \cref{fig:Nofx}, including our results up to $\cO(\ePM^{3})$ from ref.~\cite{Bautista:2026qse}.

\subsection{Matching to BHPT}
To complete the matching to the BHPT result, \cref{eq:3LoopNMatrix} must be projected onto spherical modes according to \cref{eq:deltalWQFT}, and the resulting modes must then be compared with the scattering phase shift using \cref{eq:NToPhaseShift}.
At this order, the latter is \cite{Mano:1996mf,Bautista:2023sdf}
\begin{align}
  {}_{-2}\delta_{\ell 2}^{P,(4)}
  &= \frac{\pi\epsilon_{\text{PM}}^4}{16}\,
  \frac{\mathcal{P}_4(L)}{\mathcal{D}_\ell},\label{eq:delta4}
\end{align}
where
\begin{align*}
  \mathcal{D}_\ell
  &=(2\ell+1)^3(L-2)(4L-15)(4L-3)^3 L^3,\\
  \mathcal{P}_4(L)&= 18480L^8 - 61320L^7 + 2415L^6 \\
  &\quad - 85775L^5-123233L^4 -51522L^3\\
  &\quad +953424L^2 -102816L -51840,
\end{align*}
and $L\equiv\ell(\ell+1)$.

For this purpose, it is convenient to collect the worldline result in the form
\begin{equation}
  N^{(4)}_{1}(x) = N^{(4)}_{1,\text{n.e.}}(x)+N^{(4)}_{1,\text{e.}}(x)\,,
\end{equation}
where the first term captures the non-elliptic contributions in the first two lines of \cref{eq:3LoopNMatrix};
analytically projecting this sector onto spherical modes is straightforward.

Projecting the elliptic sector is more involved, but we present here a method for doing so.
Towards this end, we rearrange the third line of \cref{eq:3LoopNMatrix} as
\begin{equation}
\begin{aligned}
  N^{(4)}_{1,\text{e.}}&=\frac{2 x \left(17805 x^4+113738 x^2+44273\right)}{1536 \pi ^2\left(1-x^2\right)}q'(x) \\
  &-\frac{\left(25485 x^4+120355 x^2+29976\right)}{1536 \pi ^2\left(1-x^2\right)}q(x)\,,\label{eq:ellip}
\end{aligned}
\end{equation}
written in terms of
\begin{equation}
  q(x) \equiv K(x^2)\mathcal{I}_2(x)\,.
\end{equation}
Concretely, our task is to perform the angular integral
\begin{align}
  \Pi_{{\rm e.},\ell}\equiv\int{\rm d}\Omega\,\, {}_2 Y_{\ell, -2}(\theta,\phi) e^{2i\phi}N^{(4)}_{1,{\rm e.}}(x).
\end{align}
Integration over \(\phi\) is trivial since ${}_{2}Y_{\ell,-2}(\theta,\phi)=e^{-2i\phi}{}_{2}Y_{\ell,-2}(\theta,0)$.

For the remaining integral it is convenient to change variables to \(x\), in terms of which the measure becomes
$
  \sin\theta\,{\rm d}\theta = 4x\,{\rm d}x .
$
In this variable, for fixed \(\ell\), the relevant
spin-weighted spherical harmonics can be written in the form\footnote{This is a consequence of the closed form \cite{Goldberg:1966uu}
\begin{equation*}
  \begin{split}
    {}_2 Y_{\ell, -2}(2 \sin ^{-1}(x),0)=&\frac{\sqrt{2 \ell+1} \left(1-x^2\right)^\ell \Gamma (\ell-1) \Gamma (\ell+3) }{2 \sqrt{\pi } \sqrt{\Gamma (\ell-1)^2 \Gamma (\ell+3)^2}}\\
    & \times\,{}_2F_1\left(-\ell-2,2-\ell;1;\frac{x^2}{x^2-1}\right)\,,
  \end{split}
\end{equation*}
for  $\ell\ge2$.
The hypergeometric series is $\sum_{k=0}^{\ell-2} c_k (x^2/(-1 + x^2))^k$.
Then, multiplying this sum by $(1-x^2)^{\ell-2}$, each term becomes of the form $c_k x^{2 k}(1 - x^2)^{\ell - 2 - k}$.
As $0\le k\le \ell-2$, this is always a polynomial in $x$.} 
\begin{equation}\label{eq:spheroidal}
  {}_2 Y_{\ell, -2}(2 \sin ^{-1}(x),0)=(1 - x^2)^2 P_{\ell}(x)\,,
\end{equation}
where \(P_{\ell}(x)\) is a polynomial in \(x\) whose degree
depends on the value of \(\ell\). 
The factor \((1-x^2)^2\) crucially brings the projection to the form
  \begin{equation}
    \Pi_{{\rm e.},\ell}=\int_0^1 {\rm d}x\,[A_{\ell}(x)q(x)+B_{\ell}(x)q'(x)]\,,
  \end{equation} 
where both \(A_{\ell}(x)\) and \(B_{\ell}(x)\) are polynomials in \(x\), easily identified by combining \cref{eq:ellip} and \cref{eq:spheroidal}, and keeping in mind the factor of $4x$ from the measure.

One can verify that \(B_{\ell}(0)=B_{\ell}(1)=0\) while $q(0)$ and $q(1)$ are finite, conveniently enabling the integration by parts of the $q'(x)$ term in the integrand without introducing a boundary
contribution:
\begin{equation}\label{eq:partialR}
    \Pi_{{\rm e.},\ell}=\int_0^1 {\rm d}x\,R_{\ell}(x)q(x),\,\,\, R_{\ell}(x)=A_{\ell}(x)-B_{\ell}'(x)\,.
  \end{equation} 
Now, $q(x)$ safisfies the Picard-Fuchs equation 
  \begin{equation}\label{eq:picard-fuchs-q}
    \mathcal{L}(q(x)) =-1\,,
  \end{equation}
where $\mathcal{L}$ is the self-adjoint operator defined in \cref{eq:picard_fuchs}.
Suppose the polynomial $R_{\ell}(x)$ is the image of some other polynomial $v_{\ell}(x)$ under $\mathcal{L}$,
\begin{equation}\label{eq:v-equation}
  \mathcal{L}(v_{\ell}(x))=R_{\ell}(x)\,.
\end{equation}
Then \cref{eq:partialR} would become 
\begin{equation}
\begin{aligned}
  \Pi_{{\rm e.},\ell}=\int_0^1 {\rm d}x\,\mathcal{L}(v_{\ell}(x))q(x) &=  \int_0^1 {\rm d}x\,v_{\ell}(x) \mathcal{L}(q(x)) \\
  &=-  \int_0^1 {\rm d}x\,v_{\ell}(x) \,,
\end{aligned}
\end{equation}
where in the first equality we used the self-adjointness of $\mathcal{L}$, and in the second, we have used \cref{eq:picard-fuchs-q}.
In the first equality we also dropped the boundary term
\begin{equation}
  \left.
    x(1-x^2)\bigl(q\,v_{\ell}' - v_{\ell}\,q'\bigr)
  \right|_{0}^{1}\,,
\end{equation}
as it vanishes.\footnote{ Indeed, as \(v_{\ell}(x)\) is a polynomial, it is
regular at both endpoints. Near \(x=0\), both \(q(x)\) and \(q'(x)\) are
finite, so the prefactor \(x(1-x^2)\) makes the boundary contribution
vanish. Near \(x=1\), \(q(x)\) is again finite. Moreover,
\begin{equation*}
  x(1-x^2) q'(x)
  =
  \left(
    \frac{E(x^2)}{K(x^2)} - (1-x^2)
  \right) q(x)
  - \frac{\mathcal{I}_1(x)}{K(x^2)} ,
\end{equation*}
 tends to zero as \(x\to1\). Hence the boundary term vanishes at both
endpoints.}

In this way, the partial wave projection of the elliptic sector is reduced to a simple polynomial integration.
It is straightforward to determine a suitable function \(v_{\ell}(x)\) for any $\ell$ by writing a polynomial ansatz, applying the operator \(\mathcal{L}\), and fixing the coefficients of the ansatz by imposing \cref{eq:v-equation}.
Incorporating the non-elliptic sector, we have confirmed that \cref{eq:NToPhaseShift} holds up to \(\ell=30\) in the helicity-preserving configuration.
Both the Magnusian and the BHPT phase shift vanish at this order when the helicity is reversed by the scattering.

We conclude that the point-particle approximation accurately describes the scattering of a gravitational wave off of a Schwarzschild black hole up to $\cO(\epsilon_{\rm PM}^{4})$.


\subsection{Analysis of the scattering amplitude}

Using integrals with only retarded propagators in our integrand produces the $\cO(\ePM^4)$ scattering amplitude, as defined by \cref{eq:DefTMatrixElement}, rather than the Magnusian.
The structure of this amplitude is more intricate than that of the Magnusian at all but the leading perturbative order.
First, $\cM_{\sigma}$ has infrared poles and the associated regularization scheme dependence.
Second, it is generally complex, while the Magnusian is real.
Third, the amplitude involves iterated integrals of weight three, while the maximum weight in the Magnusian is two.


The infrared structure of the amplitude provides important consistency checks on our calculation.
Specifically, Weinberg's soft theorem dictates that its infrared poles in $\varepsilon$ must exponentiate \cite{Weinberg:1965nx},
\begin{align}\label{eq:weinberg}
	i\cM_{\sigma}=\exp(-i\epsilon_{\rm PM}/\varepsilon)i\cM_{\sigma,\varepsilon^0}\ .
\end{align}
Here, $\cM_{\sigma,\varepsilon^0}$ starts at $\cO(\varepsilon^0)$, manifesting that all poles of the amplitude $\cM_\sigma$ come from the exponential on the right-hand side.
Thus it is immediately clear that the $\cO(\ePM^n)$ contribution to $\cM_\sigma$ has infrared poles as deep as $\vareps^{-n+1}$.
Importantly, eq.~\eqref{eq:weinberg} should be taken as an analytic formula in $\vareps$, rather than a limit, such that otherwise-vanishing $\cO(\vareps^{n\geq1})$ parts of the amplitude are included.
In particular, the interference of $\varepsilon/\varepsilon$ terms is an essential mechanism in the exponentiation of infrared poles.

To derive the perturbative version of \cref{eq:weinberg}, we expand $\cM_\sigma$ as
\begin{align}
  \cM_\sigma = \sum_{n=1}^4\sum_{m=-n+1}^{\infty}
  \ePM^n
  \vareps^m
  \cM^{(n,m)}_\sigma
  \ .
\end{align}
Similarly decomposing $\cM_{\sigma,\varepsilon^0}$ -- noting that $\cM_{\sigma,\varepsilon^{0}}^{(n,m)}=0$ for $m<0$ -- and solving \cref{eq:weinberg} perturbatively, we determine that the poles of the amplitude up to 4PM must obey
\begin{align}\label{eq:poleRelation1}
  \cM_\sigma^{(n,-m)}
  =
  \frac{(-i)^m}{m!} \cM^{(n-m,0)}_\sigma
  \ ,
\end{align}
for all pairs $(n,-m)$ except $(4,-1)$.
In this exceptional case, $\varepsilon/\varepsilon$ cancellations enter for the first time, such that \cref{eq:weinberg} demands
\begin{align}\label{eq:poleRelation}
  \cM_\sigma^{(4,-1)}
  =
  \frac12 \cM_\sigma^{(2,1)}
  -
  i\cM^{(3,0)}_{\sigma}
  \ .
\end{align}
Checking this formula necessitates the computation of $\cM^{(2,1)}_\sigma$.
Producing this missing piece, our amplitude passes the non-trivial checks in both \cref{eq:poleRelation1,eq:poleRelation}.

The finite amplitude $\cM_{\sigma,\rm fin.}$ offers an alternative route to matching with BHPT through \cref{eq:MToBHPT} \cite{Bjerrum-Bohr:2026fhs}.
It is easiest to define by inverting \cref{eq:weinberg}:
\begin{align}
  i\cM_{\sigma,\rm fin.}=\lim_{\varepsilon\rightarrow0}\exp(i\ePM/\varepsilon)i\cM_{\sigma}
  =
  \lim_{\vareps\to0}
  i\cM_{\sigma,\vareps^0}
  \ .
\end{align}
Like \cref{eq:poleRelation}, then, this finite amplitude depends on $\cO(\varepsilon^{n\geq1})$ parts of $\cM_\sigma$.
In particular, at 4PM,
\begin{align}\label{eq:4PMFiniteAmplitude}
  \cM_{\sigma,\rm fin.}^{(4)}
  =
  \cM^{(4,0)}_\sigma
  +
  i
  \cM^{(3,1)}_\sigma
  -\frac12 \cM^{(2,2)}_\sigma
  \ .
\end{align}
As the pole relations \cref{eq:poleRelation1,eq:poleRelation} provide checks on the divergent part of the amplitude, this relation can be used to check the finite part once we input data about its divergence in the forward limit.

The finite amplitude diverges quadratically as $x\rightarrow0$ at arbitrary PM orders, a behavior which is closely tied to the amplitude's infrared divergences.
Similarly to the Weinberg phase, these poles can be resummed into a so-called Newtonian phase \cite{Futterman_Handler_Matzner_1988,Dolan:2007ut,Lippstreu:2025jit}.
Following e.g. ref.~\cite{Bjerrum-Bohr:2026fhs},
\begin{align}
  \cM_{\sigma,\rm Newt.}
  =
  \ePM
  \bigg(\frac{2\omega x}{\mu e^{\gamma_{\rm E}}}\bigg)^{2i\ePM}
  \frac{\Gamma(1-i\ePM)}{\Gamma(1+i\ePM)}
  \cM_\sigma^{(1,0)}
  \ ,
\end{align}
with $\mu$ the infrared scale introduced by continuing the four-dimensional $G$ to arbitrary dimensions $G_d=G \tilde\mu^{2\varepsilon}$.
As this encapsulates the quadratic divergence, the difference
\begin{align}
  \cM_{\sigma,\rm fin.}-\cM_{\sigma,\rm Newt.}
\end{align}
should diverge at worst as $1/x$ in the forward limit.
Computing $\cM_{\sigma,\rm fin.}^{(4)}$, we confirm that this is the case.


\section{Conclusion}

We have taken the next step in the program of matching WQFT to BHPT.
Building on the methods of ref.~\cite{Bautista:2026qse}, we have demonstrated that the former formalism reproduces the phase shift for gravitational wave scattering from the latter now up to $\cO(G^{4})$.
This confirms that the point-particle approximation of Schwarzschild black holes is accurate at least up to this perturbative order.

In service of this matching, we have shown that the Murua coefficients introduced in refs.~\cite{Kim:2024svw,Brandhuber:2025igz,Gonzo:2026yha} provide an immensely economic approach for the extraction of the Magnusian.
The entirety of the cut subtraction employed in ref.~\cite{Bautista:2026qse} results in the simple Murua decomposition of master integrals in \cref{eq:MuruaIntegrals}, such that directly employing the latter alleviates the computational burden of constructing cuts explicitly.

Two remarks are in order about the function space of the amplitude and Magnusian.
First, while iterated integrals of weight three make their way into the amplitude, only weight-two and below integrals appear in the Magnusian, the same maximum weight as the previous loop order.
Interestingly, a similar phenomenon occurs at one loop, where the function space in the Magnusian is the same as tree level (polynomial), though the amplitude contains logarithms.
We might therefore expect that the maximum weight of iterated integrals appearing in the Magnusian at $\cO(G^{n})$ is $2\lfloor (n-1)/2\rfloor$.

Second, elliptic integrals have made their first appearance at this order.
At the same PM order in the two-body problem, these functions appear quadratically.
Complete elliptic integrals are periods of elliptic curves, while their squares are periods of the more complicated K3 geometry \cite{Klemm:2024wtd};
preliminary analysis indicates that the function space for four-loop gravitational wave scattering will be identical to three loops, pointing to the relative simplicity of this amplitude.
Moreover, the function space to three loops can be determined simply by analyzing the ladder topologies, begging the question whether all-loop statements about the function space can be made.

We highlight that the Magnusian continues to present an infrared-safe description of this process.
Moreover, the data up to three-loop order suggests that the worst behavior of the Magnusian in the forward limit is at tree level, with the $\cO(G^{n})$ contributions behaving as
\begin{align}\label{eq:divF}
  N\overset{x\rightarrow0}{\longrightarrow}\frac{G}{x^{2}}+\frac{G^{2}}{x}+G^{3}\log(x)+G^{4}+\cO(G^{5}).
\end{align}
It would be interesting to draw an explicit connection between these two observations, as well as to the Newtonian phase encoding the long-distance behavior of gravity \cite{Dolan:2007ut,Lippstreu:2025jit,Bjerrum-Bohr:2026fhs}.
More broadly, the infrared properties of the Magnusian demand further exploration.

Immediate next steps to this computation are clear.
Non-minimal operators in the spinless WQFT action first become relevant at the next perturbative order, such that their values can be fixed by matching the amplitude or Magnusian at $\cO(G^{5})$ to BHPT.
Additionally, it is conceptually straightforward to extend this calculation to include spin through the bosonic oscillator formalism of ref.~\cite{Haddad:2024ebn}.
As each spin order is commensurate with a PM order, the four-loop spinless amplitude in conjunction with spin-augmented results up to three loops is sufficient to match effective descriptions of both Schwarzschild and Kerr black holes to BHPT up to $\cO(G^{5})$.

\subsection*{Acknowledgments}

We thank Benjamin Sauer, and Johann Usovitsch for helpful discussions. We are particularly grateful to Gustav Mogull for insights regarding the theory of the Magnusian and the associated Murua coefficients.
This work has made use of the Black Hole Perturbation Toolkit \cite{BHPToolkit} and \texttt{FeynCalc} \cite{Mertig:1990an,Shtabovenko:2016sxi,Shtabovenko:2020gxv,Shtabovenko:2023idz}.
Integrand generation and tensor reduction was done with \texttt{FORM} \cite{Ruijl:2017dtg}. IBP reduction was performed with \texttt{Kira 3} \cite{Lange:2025fba}.
Canonicalization was performed with the help of \texttt{CANONICA} \cite{Meyer:2017joq} and \texttt{INITIAL} \cite{Dlapa:2022wdu}.
Numerical integration was performed with \texttt{AMFlow} \cite{Liu:2022chg}; we are immensely grateful to Xiao Liu, Benjamin Sauer, and Johann Usovitsch for assistance in implementing this package.
Finally, we thank Giacomo Brunello, Mario Meo, and Sid Smith for coordinating the realease of their work \cite{Brunello:2026rdk}.
AI tools were used for coding and research.
The work of M.D. and K.H.~was funded by the European Union through the 
European Research Council under ERC Advanced Grant 101097219 (GraWFTy). The work of
Y.F.B. has been supported by the European Research Council under Advanced Investigator Grant ERC–AdG–101200505.
Views and opinions expressed are however those of the authors only and do not necessarily reflect those of the European Union or European Research Council Executive Agency. Neither the European Union nor the granting authority can be held responsible for them.

\appendix

\bibliographystyle{apsrev4-2}
\bibliography{ThreeLoopCompton}

\end{document}